\documentclass{article}

\usepackage{PRIMEarxiv}

\usepackage[utf8]{inputenc} 
\usepackage[T1]{fontenc}    
\usepackage{hyperref}       
\usepackage{url}            
\usepackage{booktabs}       
\usepackage{amsfonts}       
\usepackage{nicefrac}       
\usepackage{microtype}      
\usepackage{lipsum}
\usepackage{fancyhdr}       
\usepackage{graphicx}       
\graphicspath{{media/}}     

\title{Taming the Ransomware Threats: Leveraging Prospect Theory for Rational Payment Decisions
\thanks{\textit{\underline{Citation}}: 
\textbf{Authors. Title. Pages.... DOI:000000/11111.}} 
}

\author{
  Pranjal Sharma \\
  Department of Computer Science\\
  Heidelberg University \\
  Heidelberg, Germany\\
  \texttt {pranjaldub@gmail.com}
}

\begin{document}
\maketitle

\begin{abstract}
Day by day, the frequency of ransomware attacks on organizations is experiencing a significant surge. High-profile incidents involving major entities like Las Vegas giants MGM Resorts, Caesar Entertainment, and Boeing underscore the profound impact, posing substantial business barriers. When a sudden cyberattack occurs, organizations often find themselves at a loss, with a looming countdown to pay the ransom, leading to a cascade of impromptu and unfavourable decisions.
This paper adopts a novel approach, leveraging Prospect Theory, to elucidate the tactics employed by cyber attackers to entice organizations into paying the ransom. Furthermore, it introduces an algorithm based on Prospect Theory and an Attack Recovery Plan, enabling organizations to make informed decisions on whether to consent to the ransom demands or resist. This algorithm Ransomware Risk Analysis and Decision Support (RADS) uses Prospect Theory to re-instantiate the shifted reference manipulated as perceived gains by attackers and adjusts for the framing effect created due to time urgency. Additionally, leveraging application criticality and incorporating Prospect Theory's insights into under/over weighing probabilities, RADS facilitates informed decision-making that transcends the simplistic framework of "consent" or "resistance," enabling organizations to achieve optimal decisions.
\end{abstract}


\section{Introduction}

\begin{figure}
\centering
\includegraphics[width=15 cm]{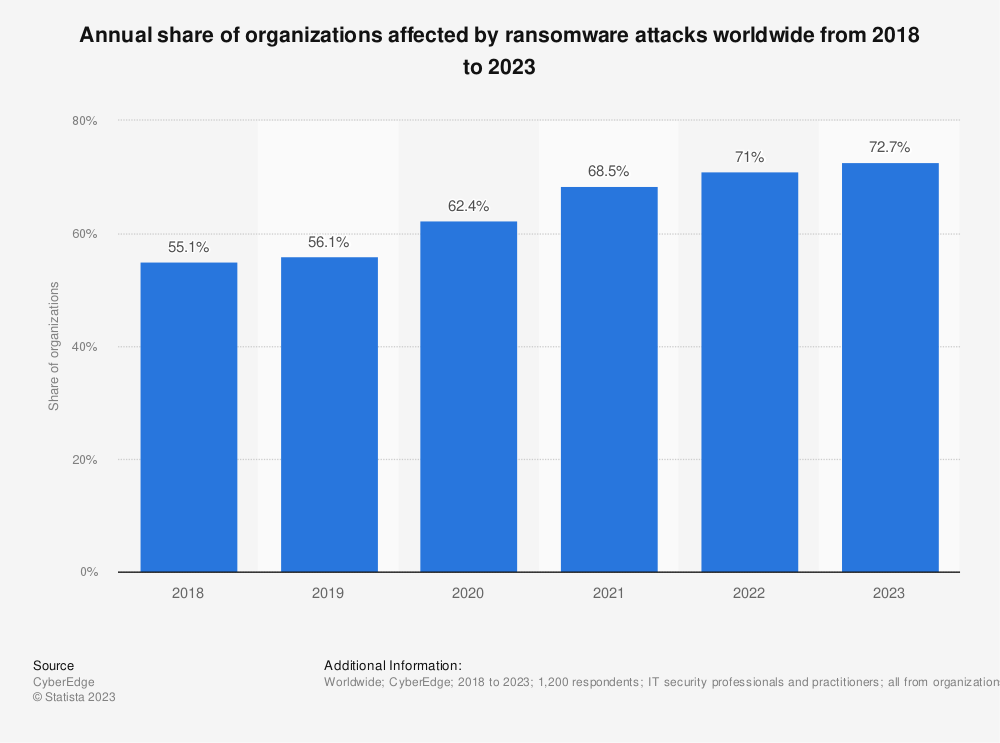}
\caption{Statistics showing growth in ransomware attacks worldwide between the year 2018-2023\label{fig1}}
\end{figure}

The world is elevating new technologies in day-to-day life making life easier. Entertainment, Transportation, Hospitality, and Medicine are such good examples of innovation centres, but these new technologies come with new security flaws. According to statistical research conducted by Statista, there has been a noteworthy 17\% increase in cyberattacks based on ransom over the past five years \ref{fig1} \cite{bib_statista}. Every year between 3000-6000 security flaws (CVEs) are reported, with countless more incidents arising from the exploitation of these vulnerabilities, often resulting in ransom demands \cite{bib_cvedetails}.

In the recent cyber attack on MGM Resorts and Caesar Entertainment (10 Sep'2023), where ATM and Slot machines were not working for dispensing cash, digital key cards could not open the hotel rooms, not accepting credit cards for payments, online reservations not available or not confirmed, TV service/phone line not working in hotel rooms, cash only transaction, pen and paper used for some transactions. This not only created havoc for both but also the share price was approx. 22\% down. Additionally, businesses were losing millions every day. In the end, Caesar paid 10.6M USD while MGM didn't pay and tried to recover the business on its own. In this duration of decision, the organisation lost a lot of business which could have been quick if there had been some automated process or tool to accumulate all the values around these kinds of attacks \cite{bib_techcrunch}.

Now we build a scenario of a ransom attack on some random organization to understand the values and responsible person behind this intrusion, and further try to fetch data for our RADS algorithm. What happens just after the cyber-attack? There comes the urgency of making the decision. What decisions are required to make in such a breach? This question will be answered later time as we endeavour to grasp the baseline behind these decisions.

\subsection{ISO 27001:2022}
According to ISO 27001:2022 Annex A.16 Information Security Incident Management \cite{bib_iso}, there are various steps required to be part of controlling security incidents. Some of those steps are as follows:
\begin{enumerate}
    \item A.16.1.1 Responsibilities and Procedures
    \item A.16.1.2 Report Security Events
    \item A.16.1.4 Assessment of Decision on Information Security Events
    \item A.16.1.5 Response to Information Security Incidents
    \item A.16.1.7 Collection of Evidence          
\end{enumerate}
There is another fact that needs to be understood from ISO27001:2022 i.e. Annex A.17 Information security aspects of Business continuity management [4]. Some notions required here are:
\begin{enumerate}
    \item A.17.1.1 Planning Information Security Continuity
    \item A.17.1.2 Implementing Information Security Continuity
\end{enumerate}
What is the information that we are looking here for in these ISO standards for our purpose?
The paper here presents the baseline that is required for our tool to be implemented. We require these baselines to be already enabled to understand the current decision process after a digital breach. If we investigate Annex A.16.1.1, there must be a responsible person who handles these security events and procedures to understand the steps required to implement a fix to the issue and to brief Annex A.17.1.2 there must be a plan implemented to restore the business after the ransom attack.

\subsection{Cyber Attack Process}
Now we describe the decisions required after a security incident according to the above information and reform them into steps as follows:
\begin{enumerate}
    \item After a cyber-attack, a security incident is created by the application owner, multiple incidents for different applications.
    \item Ransom attacker initiates a countdown for ransom payment while notifying the organization of the breach.
    \item Attacker also declares that after the conclusion of the countdown, the ransom amount will be doubled.
    \item Impact is analyzed for this breach by application owners.
    \item Cost and Time estimation of the data breached.
    \item Value of ransom demanded by the attacker.
    \item Feasibility to restore the complete data.
    \item In real scenarios, it is feasible to restore the systems due to recent backup still small current data is still lost.
    \item The value of current data is of utmost importance. For example, banking data, stock market etc. therefore, to return at the exact current situation is relatively less feasible.
    \item Business compares the cost of the data breached, ransom value, and feasibility. Additional factors such as legality, certainty and assurance are also looked into.
    \item Business decides whether to pay the ransom or not based on the above factors.
\end{enumerate}

Some questions can be posed based on the above steps. What is new in the above steps defined? Are values like feasibility, impact, etc. are the deciding factors?

\subsection{Factors Affecting Ransom Payment \cite{bib_pwc}}

Following a cyber-attack, there are some factors required to be evaluated to conclude a decision if the organization should pay the ransom or try to restore the systems itself, and risks must be weighed properly. Factors that can be decisive in this decision-making process:

\begin{enumerate}
    \item Feasibility: Are backups or other means available to restore the business?
    \item Effort: What are the overall Time and Cost efforts to restore the business?
    \item Impact: What would be the impact of delays due to the restoration process on the business?
    \item Legality:  What legal risks are the organization exposed to?
    \item Assurance: What is the assurance that the data will be recovered after paying the ransom?
    \item Certainty: Is there are certainty that the ex-filtrated data will not be passed on or sold to third parties?
    \item Cyber-criminal Funding: Ransom payments also fund the continued activity of cybercriminals.
\end{enumerate}

\subsection{Prospect Theory}
If taken look at the above process step 2 and step 3 have some psychological benefits for the attackers. According to the psychological decision-making framework Prospect theory developed by Daniel Kahneman and Amos Tversky in 1979 states that people are risk-averse in the domain of gains and risk-seeking in the domain of losses \cite{bib_prospecttheory}. Prospect Theory can be described via three following points:
\begin{enumerate}
    \item Loss Aversion - According to this, the pain of losing is psychologically twice as much as the pleasure of gaining.
    \item Reference Dependence and Framing Effect: According to reference dependence people make decisions not in absolute terms but about some reference point or previous reference decision. The framing effect states that the decisions can be influenced based on how the choices are framed.
    \item Non-Linear Weighing of Probabilities: A critical point of human decision-making is that they do not weigh the choices linearly. They overweight the small probabilities and underweight the high probabilities.
\end{enumerate}

\begin{figure}
\centering
\includegraphics[width=15 cm]{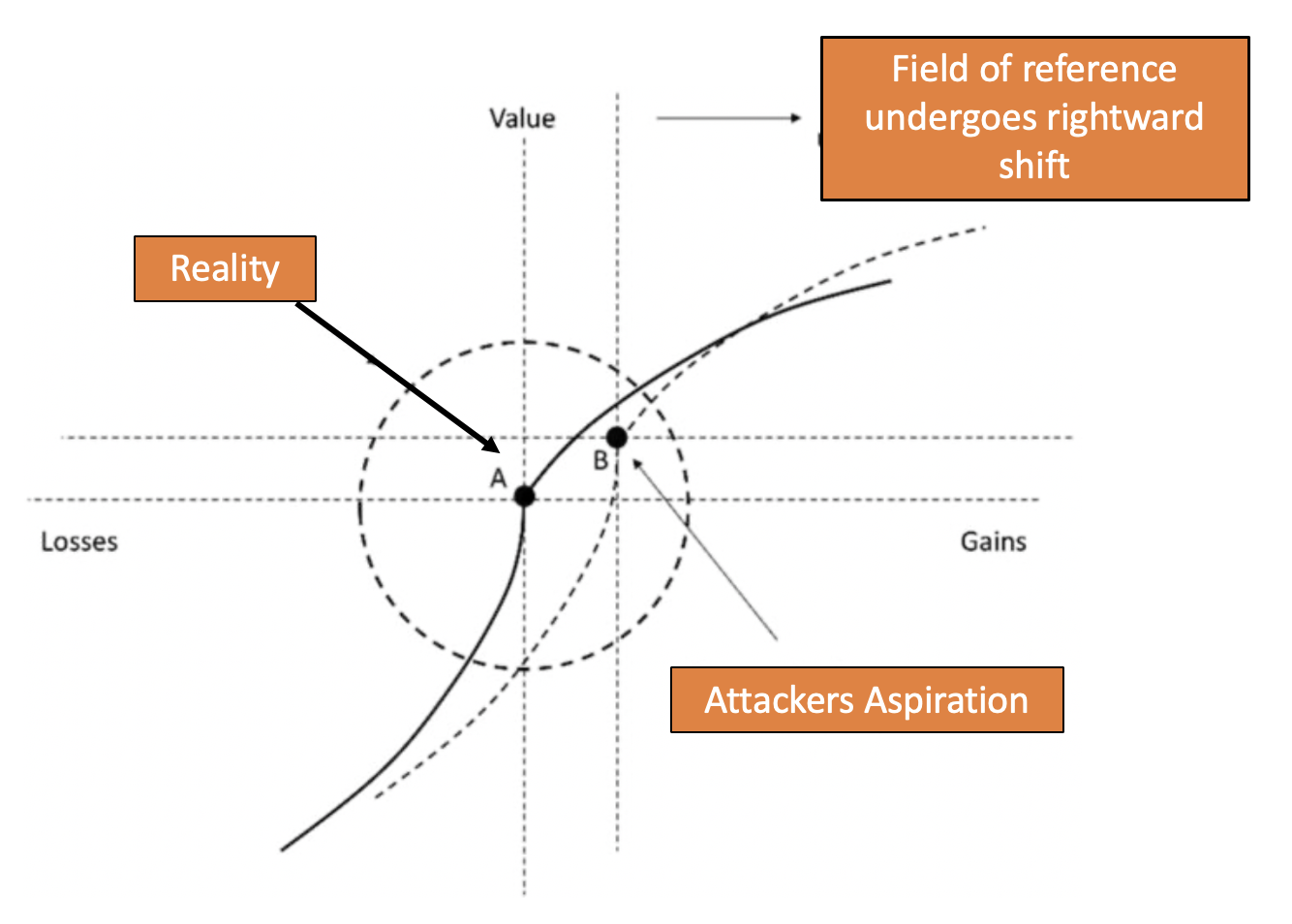}
\caption{Showing attackers' aspiration of perceived gains in the domain of losses.\label{fig2}}
\end{figure}

Here, the attackers try to manipulate the positive decision on ransom payment as perceived gains.
Also, corporations sometimes overweight/underweight certain probabilities during decision-making in the presence of high risks which certainly creates ambiguity in the s-shaped function of this cyber breach ransom payment decision as in \ref{fig2}.
We would like to adjust back the effect of various aspects of prospect theory and present a final calculation-based decision. It saves a lot of time but also reflects precise decisions on ransom payments.

This paper is divided into four important segments First, Methodology where the paper emphasizes the values that are gathered from the prospect theory which will help optimize the factors that affect ransom payment decisions. Second, the Results will help us understand the real-world scenario and how effective this algorithm would be in saving time by comparing these two decisions. Third, the Conclusion helps to summarize the effectiveness of the RADS tool we have provided, Fourth, the future work will help readers comprehend the current research and work further on improving the effectiveness of decision-making.

\section{Methodology}
\subsection{Prospect Theory}
The initial requirement poses the necessity to describe Prospect Theory and how it is being used by the attackers to shift the s-shaped function. Here, the attacker uses some points from Prospect Theory as follows:
\begin{enumerate}
    \item Reference Effect - Attackers manipulate the reference point by probating the secrets of the organisation, future cyber attacks on the organization, leaking data on the internet, and complete business failure which makes little ransom payment look like an intangible task.
    \item Framing Effect - Attackers leverage the Prospect Theory framing certain decisions in terms of losses or gains which can be elaborated as below:
    \begin{enumerate}
    \item Emphasizing Loss - The attacker lists all the data they have obtained by the ransom attack or how much data they have encrypted and rendered unusable.
    \item Urgency and Time Frame - Attackers impose deadlines such as the ransom amount will be doubled after 7 days, to create urgency and time frame.
    \item Potential Further Losses - Attackers emphasize that after the deadline they will leak the data on the internet or they will delete the key for encrypted data.
    \item Ransom Payment Framed as Gains - Attackers showcase paying ransom for the attack as gains.
    \end{enumerate}
    \item Overweighing of Probabilities - Often attackers hassle the ransom payment to take urgency advantage, wherein organizations do not have enough time to calculate the proper impact of the attack. Therefore, these victims in several instances improperly weigh the overall damage caused by the cyber attackers.
\end{enumerate}

\subsection{Proposed Solution}
\begin{enumerate}
    \item Reference Effect - It is extremely difficult to re-instantiate back to the ORIGINAL state i.e. state just before the cyber attack, but getting to a BETTER state is feasible. Here, this problem of shifted reference could be dealt with by studying the previous cyber attacks on similar organizations(size of organization). Additionally, how well did the organization do in the previous attacks(if any)? Thus, defining the value of the reference effect.
    \item Framing Effect - We can weigh the framing effect based on how much the attacker has created urgency of time.
    \item Non-linear Weighing of Probabilities - We use the criticality of the applications to linearly weigh the probabilities of the decision factor.
\end{enumerate}

\subsection{Baseline for RADS}
\begin{enumerate}
    \item Wait for the cyberattack event.
    \item Calculate the Impact \\
        Impact Data: \verb|D_Impact| \\
        Weight of Impact = \verb|W_Impact|

    \item Estimate The Feasibility \\
        Feasibility: \verb|D_Feasibility| \\
        Weight of Feasibility: \verb|W_Feasibility|
	\item Estimate Effort in terms of Time and Cost
        \begin{enumerate}
            \item Time Effort Data: \verb|D_Effort_Time| \\
                Weight of Time Effort: \verb|W_Effort_Time|
            \item Cost Effort Data: \verb|D_Effort_Cost| \\
                Weight of Cost Effort: \verb|W_Effort_Cost|
        \end{enumerate}
        
	\item Calculate Framing effect \\
        Value of Framing effect: \verb|V_Framing| \\
        Weight of Framing effect: \verb|W_Framing|
        
	\item Calculate Reference effect \\
        Value of Reference effect: \verb|V_Reference| \\
        Weight of Reference effect: \verb|W_Reference|
	\item  Ransom Demanded  \\
        Ransom Value = \verb|V_Ransom|

	\item Calculate the value of data breach \\
        Value of Data Breached = \verb|V_Breached|
\end{enumerate}

\begin{figure}
\includegraphics[width=17cm, height=20cm]{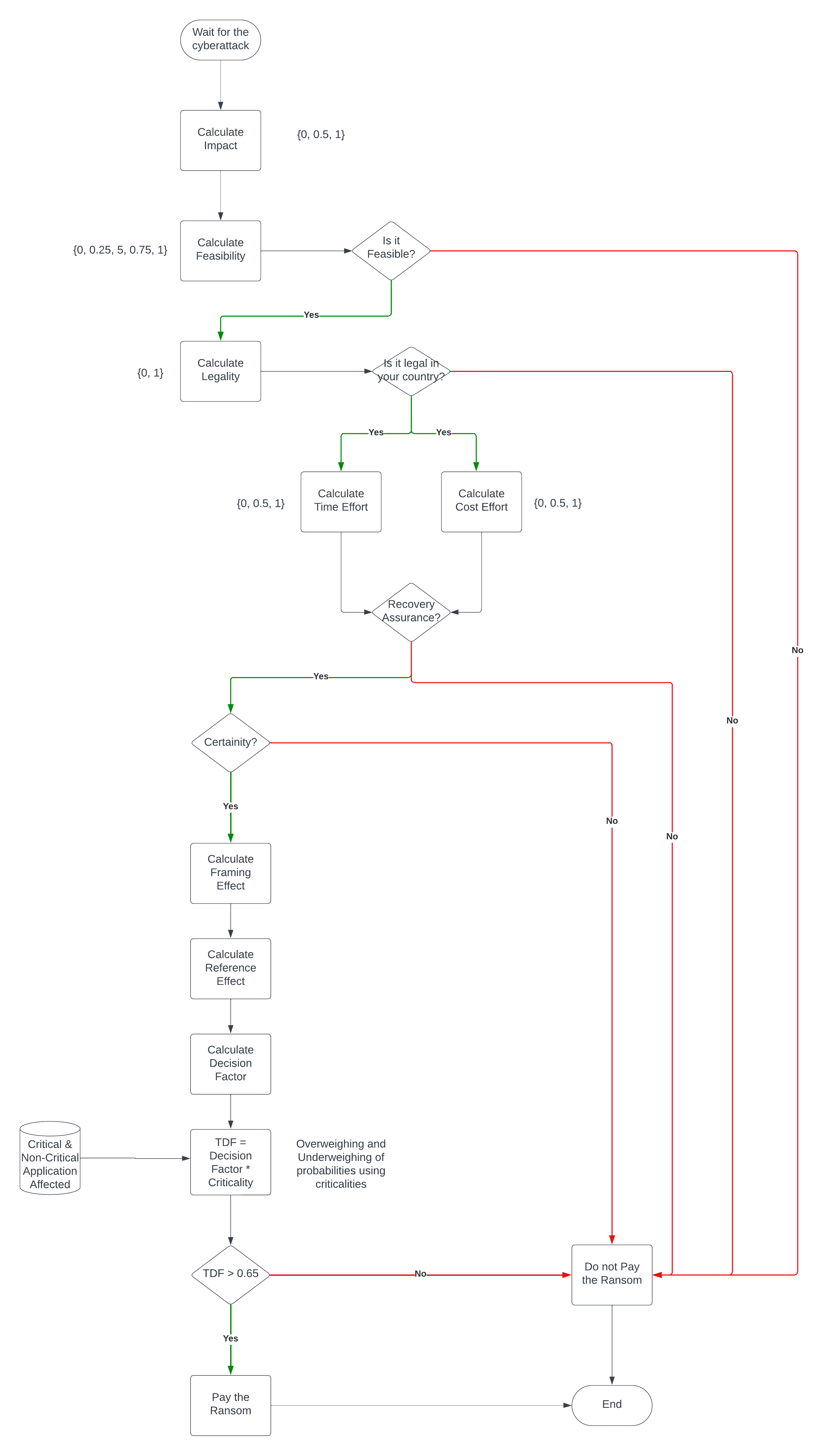}
\caption{Ransomware Risk Analysis and Decision System \label{fig3}}
\end{figure}

\subsection{Ransomware Risk Analysis and Decision Support (RADS) \ref{fig3}} 

Here, we discuss the case where the attackers have encrypted the organization's data.
\begin{enumerate}
    \item IMPACT ESTIMATION
        
        Impact Data =  \verb|D_Impact| ->  Range \{0, 0.5, 1\} \\
        0 = No Impact \\
        0.5 = Medium Impact \\
        1 = High Impact \\
        Weight of Impact = \verb|W_Impact|

	\item FEASIBILITY ESTIMATION \\
        Feasibility Data = \verb|D_Feasibility|  ->  Range \{0, 1\} \\
        0 – Not feasible at all i.e. no backups \\
        0.25 – Some data has backup but not all \\
        0.5 – at least half of the data backup is available \\
        0.75 – Most data have been backup \\
        1 – The last backup was created recently to the cyber attack \\
        Weight of Feasibility = \verb|W_Feasibility|
        
    \item EFFORT ESTIMATION \\
        Time Effort Data = \verb|D_Effort_Time| -> Range \{0, 0.5, 1\} \\
        Cost Effort Data = \verb|D_Effort_Cost| -> Range \{0, 0.5, 1\} \\
        Weight of Effort Time = \verb|W_Effort_Time| \\
        Weight of Effort Cost = \verb|W_Effort_Cost|

    \item FRAMING EFFECT \\
        Value of Framing effect = \verb|V_Framing| – This will be derived from how much the scenario has been framed based on the time duration given. Range \{0, 0.5, 1\}\\
        0 – No time limit \\
        0.5 – Time limit is approximately equal to recovery time \\
        1 – Very short period to pay the ransom \\
        Weight of Framing effect = \verb|W_Framing|

	\item REFERENCE EFFECT \\
        Value of Reference effect = \verb|V_Reference| - This will be derived from the previous cyber attacks if any and how well the organisation performed against these attacks. Range \{0, 0.5, 1\}\\
        0 – No previous incidents\\
        0.5 – Incidents occurred in the past and were treated insufficiently \\
        1 – Incidents occurred in the past and were treated sufficiently \\
        Weight of Reference effect = \verb|W_Reference|

	\item DECISION FACTOR \\
        The calculation of the decision factor is for each application affected, for n number of applications affected, calculate the Decision Factor for all the applications i.e. \\
        DF1, DF2, DF3……... DFn.\\
        \verb|DF1 =| \verb|(D_Impact * W_Impact)| + \verb|(D_Feasibility * W_Feasibility)| + \\ \verb|(D_Effort Time * W_Effort Time)| + \verb|(D_Effort Cost * W_Effort Time)| \\ - \verb|(V_Framing * W_Framing)| + \verb|(V_Reference * W_Reference)|   

	\item VALUE OF THE DATA BREACHED \\
        \verb|V_Breached|

    \item Criticality of Applications - Export the Application Criticality and assign them in the below variables. \\
        CAP1, CAP2, CAP3……… CAPn

    \item Total Decision Factor \\
        Mean Criticality - Mean of all the criticality of applications \\
        Weight Criticality - Sum of all the criticality of applications\\

    \item Normalization of Decision Factors for Critical, High and Low Application
        \begin{enumerate}
            \item \verb|Adjusted Factor Critical_i = TDF * (1 + (CAPi - Mean Criticality) / Weight Criticality)|\\
            \mbox{Where, CAPi is the Criticality of $  i^{th}  $ Critical application}
            \item   \verb|Adjusted Factor High_i = TDF * (1 + (CAP_i - Mean Criticality) / Weight Criticality)|\\
            \mbox{Where, CAPi is the Criticality of $  i^{th}  $ High application}
            \item Adjusted Factor Medium is Not required to adjust the weight of medium criticality applications
            \item \verb|Adjusted Factor Low_i = TDF * (1 - (CAP_i - Mean Criticality) / Weight Criticality)|\\
            \mbox{Where, CAPi is the Criticality of $  i^{th}  $ Low application} \\
        \end{enumerate}
    \item If the Adjusted Factor > 65;\\
    Then pay the ransom for those applications.\\
  \end{enumerate}      
\textbf{*Note Organisations have to weight each of these values in advance as per their requirements} \\
\\
\textbf{**Note Organisations can choose the Range for each of the baselines according to their needs (the above ranges are just examples to justify the baselines)} \\
\textbf{***Note Sum of Weights = 100}\\

\section{Results}

Consider an organization The Crown Financing GMBH registered in Heidelberg, Germany. The Crown GMBH have 30 applications (let). In these applications, 3 applications are critical {1}, 1 Medium Criticality {0.5} and 26 Low Criticality {0.25} applications. Backups are taken weekly on Mondays in this organization for all applications irrespective of the criticality of the application. The organization has not experienced any cyberattack in the past so it has no previous knowledge of how to completely handle such cyber attacks.

On a nice Wednesday evening, The Crown Financing GMBH experienced a ransomware attack on its assets, affecting 1 Critical application and 15 low-criticality applications. All these assets have been encrypted by a key which will be released by the attackers if the ransom is paid within 28 days. For the first 7 days, the ransom amount is 10M USD. It will double up every 7 days.

We consider the weight of each factor for this experiment to contribute 20\%

VImpact = 0.5\\
WImpact = 20\\
\\
VFeasibility = 0.75\\
WFeasibility = 20\\
\\
VEffort Time = 0.25\\
WEffort Time = 10\\
\\
VEffort Cost = 0.75\\
WEffort Cost = 10\\
\\
VFraming = 0.75\\
WFraming = 20\\
\\
VReference = 0.75\\
WReference = 20\\
\\
Decision Factor = 0.5*20 + 0.75*20 + 0.25*10 + 0.75*10 - 0.75*20 + 0.75*20 = 10 + 15 + 2.5 + 7.5 - 15 + 15 = 35

If the above factors are different for different applications we calculate it for each application otherwise DF for each application is equal.

TDF = DF1*CAP1 + DF2*CAP2 + DF3*CAP3......DF

CAP1 = 5 (1 Critical application)
CAP 2 = CAP3 = ...... CAP16 = 2 (15 Low Criticality applications)

Total number of affected applications = 16

For TDF, we need to normalize the variations:

Weight Criticality = 5 + 2 + 2.....+2 = 5 + 15 *2 = 35
Mean Criticality = 35/16 = 2.1875

TDF 1 = Adjusted Factor Critical1 = DF1 * (1 + (CAP1 - Mean Criticality) / Weight Criticality) \\
Adjusted Factor Critical1 = 35 * (1+ (5-2.1875)/35) = 35* 1+(2.8125/35) = 37.1875

TDF 2 = Adjusted Factor Low = Decision Factor * (1 - (Mean Criticality - CAP2) / Weight Criticality

Adjusted Factor Low = 35 * (1 - (2.1875 - 2)/35) = 35 * (1 - 0.1875/35) = 34.8125

Now, we can pay the ransom just for the applications that have a TDF above 65. We don't have TDF above 65 so the organisation should recover itself which will incur the least loss.

\section{Conclusion}
The rapid pace of technological advancement creates a constant arms race against those who seek to exploit vulnerabilities for personal gain. Understanding the mindset of attackers is essential for proactive defence. This paper offers a new approach to cyber security by blending technical analysis with insights from behavioural economics.  Prospect Theory sheds light on the attacker's decision-making processes, while an attack recovery plan outlines potential difficulties while deciding to pay ransom. The RADS algorithm acts as a powerful decision-support tool, allowing us to simulate how attackers strategically exploit psychological factors to manipulate decision-making in their favour. Our research aims to provide critical decisions in the form of an algorithm, leading to optimal ransom payment decisions.

\acknowledgments{The authors of this paper acknowledge the formal feedback from Prof. Dr Rima-Maria Rahal from the Department of Psychology, Heidelberg University.}

\conflictsofinterest{``The authors declare no conflicts of interest.'' ``The funders had no role in the design of the study; in the collection, analyses, or interpretation of data; in the writing of the manuscript; or in the decision to publish the results''.} 



\abbreviations{Abbreviations}{
The following abbreviations are used in this manuscript:\\

\noindent 
\begin{tabular}{@{}ll}
RADS & Ransomware Risk Analysis and Decision Support\\
GMBH & “Gesellschaft mit beschränkter Haftung,” translates to “company with limited liability.”\\
ISO & International Organization for Standardization\\
CVE & Common Vulnerabilities and Exposures\\
POV & Point of view
\end{tabular}
}

\bibliographystyle{unsrt}  
\bibliography{references}

\end{document}